\documentclass[iop]{emulateapj}
\usepackage{graphicx}
\usepackage{mathtools}
 \usepackage[applemac]{inputenc}
 \usepackage[usenames]{color}	
\usepackage{amsmath}
\usepackage{mathrsfs}
\usepackage{dsfont}
\usepackage{amssymb}
\usepackage{mathtools}
 \usepackage[applemac]{inputenc}
\usepackage{hyperref}
\usepackage{soul}
\usepackage{footmisc}
\usepackage{booktabs}
\usepackage{natbib}
\usepackage{url}
\usepackage[dvipsnames]{xcolor}

\usepackage{epsfig}
\usepackage{subfigure}
\usepackage{multirow}

\newcommand{\red}{\textcolor{black} }

\shorttitle{Lorentzian asymmetry: Frequency biases}
\shortauthors{Benomar et al.}

\begin{document}


\title{Asymmetry of line profiles of stellar oscillations measured by Kepler for ensembles of solar-like oscillators: impact on mode frequencies and dependence on effective temperature}

\author{O. Benomar\altaffilmark{1}, Mjo. Goupil\altaffilmark{2}, K. Belkacem\altaffilmark{2}, T. Appourchaux\altaffilmark{3}, M.B. Nielsen\altaffilmark{1}, 
	M. Bazot\altaffilmark{1}, L. Gizon\altaffilmark{1,4,5}, S. Hanasoge\altaffilmark{1,6}, K.R. Sreenivasan\altaffilmark{1,7}, \red{B. Marchand}\altaffilmark{1}} 

\altaffiltext{1}{NYUAD Institute, Center for Space Science, New York University Abu Dhabi, PO Box 129188, Abu Dhabi, UAE} 
\altaffiltext{2}{LESIA, UMR8109, Universit\'e Pierre et Marie Curie, Universit\'e Denis Diderot, Obs. de Paris, 92195, Meudon Cedex}
\altaffiltext{3}{Univ. Paris-Sud, Institut d'Astrophysique Spatiale, UMR 8617, CNRS, B\^atiment 121, 91405, Orsay Cedex}
\altaffiltext{4}{Max-Planck-Institut f\"ur Sonnensystemforschung, 37077 G\"ottingen, Germany}
\altaffiltext{5}{Institut f\"ur Astrophysik, Georg-August-Universit\"at G\"ottingen, 37077 G\"ottingen, Germany}
\altaffiltext{6}{Tata Institute of Fundamental Research, Mumbai, India 400005}
\altaffiltext{7}{New York University, NY 10012, USA}

\begin{abstract}
Oscillation properties are usually measured by fitting symmetric Lorentzian profiles to the power spectr\red{a} of Sun-like stars. 
However the line profiles of solar oscillations have been observed to be asymmetrical for the Sun. The physical origin of this line asymmetry is not fully understood, although it \red{should}  depend on the depth dependence of the source of wave excitation (convective turbulence) and details of the observable (velocity or intensity).
For oscillations of the Sun, it has been shown that neglecting the asymmetry leads to systematic errors in the frequency determination. This \red{could} subsequently affects the results of seismic inferences of the solar internal structure.  
%
%
%
Using light curves from the {\it Kepler} spacecraft we \red{have} measured mode asymmetries in 43 stars. We confirm that neglecting the asymmetry leads to systematic errors that can exceed the $1\sigma$ confidence intervals for seismic observations longer than one year. Therefore, the application of an asymmetric Lorentzian profile is to be favoured to improve the accuracy of the internal stellar structure and stellar fundamental parameters. 
We also show that the asymmetry changes sign between cool Sun-like stars and hotter stars. This provides the best constraints to date on the location of the excitation sources across the Hertzsprung-Russel diagram. 
\end{abstract}


\section{Introduction}\label{sec:1}
Spaceborne instruments such as MOST \citep{Walker2003}, CoRoT \citep{Baglin2006a} and {\it Kepler} \citep{Borucki2010} initiated an era of high-precision asteroseismic measurements.
In particular with {\it Kepler}, the level of resolution achieved for stellar pulsation frequencies is remarkable (few tens of nano-Hertz) 
and has lead to a measurement precision equivalent to helioseismic measurements from a decade ago\red{, despite the distances to these stars compared to the Sun}. This has enabled us to reach unprecedented precision in the determination of both the fundamental parameters of stars (such as their mass, radius, and age) as well as their internal structure. However, many sources of systematic error remain that need to be assessed and mitigated.

Seismic inferences of stellar interiors and fundamental parameters of low-mass stars rely on pulsations frequencies measured from the power spectra of its light curve \citep[e.g.][]{Davies2015, Benomar2009b,Appourchaux2008, Appourchaux1998, Chaplin2001}. This requires fitting a model to the power spectrum, \red{which in turn is} based on \red{a set of} approximations\red{. These approximations} could be a source of systematic error on pulsation frequencies, \red{which} will impact the fundamental parameters of stars derived from stellar evolution and pulsation codes \citep[e.g.][]{Paxton2013,JCD2008a,Morel2008, JCD2008b,Scuflaire2008}.

Solar-like oscillations are excited by the turbulent convection in the outermost layers of convective envelopes of low-mass stars. As a first approximation such modes can be described by stochastically damped oscillators, whose solution in the power spectrum are Lorentzian profiles \citep{Duvall1988, Libbrecht1988, Kumar1988}. Current pulsation analysis methods commonly assume that this approximation is accurate. 

However, helioseismic observations by \cite{Duvall1993} showed that the line profiles of solar oscillations are asymmetrical. Mode line profiles observed in velocity have more power on the low-frequency side of the power maximum, while line profiles observed in intensity have more power on the high-frequency side of the power maximum. 

In this framework, many studies attempted to interpret the solar mode asymmetry.
It was recognised immediately that a source of wave excitation that is localised \red{ below the surface} would lead to a line profile asymmetry \citep{Duvall1993, Gabriel1993}. A line asymmetry will result from the interference between waves coming directly from the source to the observer and waves experiencing multiple reflections inside the stellar oscillation cavity before reaching the observer. However, \red{this} theory has little predictive power as the source of wave excitation is not known in detail, and there is no universal explanation for the reversal of asymmetry between the intensity and velocity observations. The reversal of asymmetry has been attributed either to the presence of noise of convective origin that is correlated with the waves \citep{Roxburgh1997, Nigam1998b} or to radiative transfer effects \citep{Georgobiani2003}.


It was found by \cite{Duvall1993} and \citet[][TT98 hereafter]{Toutain1998}, that neglecting mode asymmetry leads to systematic errors in the determination of \red{solar} mode frequencies, which \red{may} subsequently leads to imperfect modeling of the Sun. This motivated several works that \red{attempted} to properly account for mode asymmetry when fitting oscillation modes in the power spectrum of the Sun \citep[e.g.][]{Chaplin1999, Korzennik2005, Vorontsov2013}.

However, while mode asymmetry is now considered as an essential ingredient to obtain an accurate measurement of mode frequencies and then an accurate modelling of the Sun, it has been neglected so far for Sun-like stars. Our objective is thus to provide the first measurements of mode asymmetry in Sun-like stars and investigate its evolution across the HR diagram. This is made possible by using the {\it Kepler} spacecraft photometric measurement\red{s} (i.e. in intensity). 

\section{Analysis of the power spectrum with mode asymmetry} \label{sec:2}
The analysis of the pulsation modes of Sun-like stars is performed in the power spectrum and uses the Bayesian statistical algorithm of \cite{Benomar2009} with improvements and priors as defined in \cite{White2017}. We recall here the main properties of the analysis and develop in more details the specifics of the mode asymmetry measurements.
   
\subsection{Modeling of the mode asymmetry}
Several parametric expressions for asymmetrical line profiles have been proposed, see e.g. Appendix A in \citet{Vorontsov2013}. The most commonly used asymmetrical line profile is the asymmetrical Lorentzian profile proposed by \cite{Nigam1998},
\begin{equation}
\label{eq:power_Lorentzian}
L_{n,l,m}(\nu)= H_{n,l,m}  \frac{ (1 + B_{n,l,m} x )^2 + B^2_{n,l,m} }{1 + x^2}
\end{equation}
with $x=2 (\nu-\nu_{n,l,m})/ \Gamma_{n,l,m}$. Here, the variables $\nu_{n,l,m}$,  $H_{n,l,m}$ and $\Gamma_{n,l,m}$ are the mode frequency, height and the full width at half maximum (simply referred \red{to} as the mode width), respectively. $B_{n,l,m}$ is the asymmetry coefficient. 
Note that this parametrization is only meaningful close to \red{the} resonance \red{frequency of the modes, while our procedure involves a global fit of the power spectrum. Because of this, systematic errors may occur on the measured asymmetry parameter. To minimize these potential systematics, the Lorentzians are only calculated for, $|\nu - \nu_{n,l,m}| < c (l\, \nu_s + \Gamma_{n,l,m})$. Here, $c=10$ is the truncation parameter and $\nu_s$ is the rotational splitting. For a Sun-like star, this implies that the Lorentzian is only calculated  on a window of width $\approx 20\mu$Hz, centered on the central mode frequency. This may be comparable to performing a local fit of the mode, but allows us to still account for the common information among the modes (noise background, splitting $\nu_s$, stellar inclination, etc.). The accuracy of this approach is further discussed in section \ref{sec:ass_asym}.} 

This asymmetry coefficient depends on the mode width, so that \cite{Gizon2006} suggested \red{using} the dimensionless asymmetry coefficient $\chi_{n,l,m}$, 
\begin{equation}
 	\chi_{n,l,m}=2 \frac{B_{n,l,m} \, \nu_{n,l,m}}{\Gamma_{n,l,m}}, 
    \label{eq:asym_chi}
\end{equation} 

\begin{figure}[t]
  \begin{center}
  	\epsfig{figure=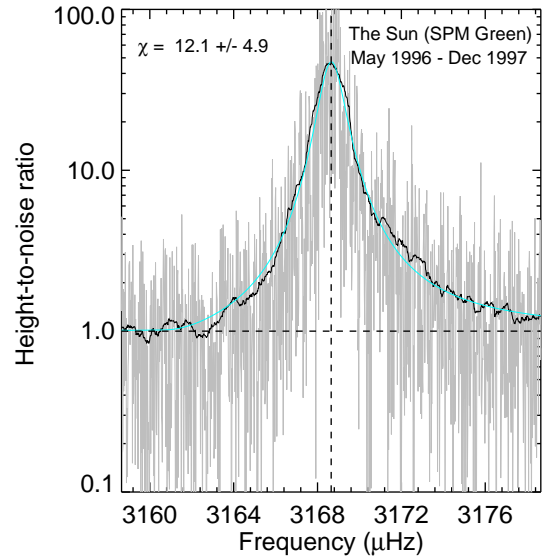, angle=0, width=7cm} 
   \end{center}
\caption{Example of mode asymmetry in the solar power spectrum for a mode of degree $l=0$. 
Gray is the original spectrum. Black is \red{the} box-car smooth\red{ed spectrum, with a kernel} of width equal to the mode width. Blue is the best fit. The Sun has a positive asymmetry $\chi$.} 
\label{fig:Asym_PSF_Sun}
\end{figure}

	 The mode asymmetry increases the power on one side of a mode, by elevating the tail of the Lorentzian. When the asymmetry is positive, the tail is raised at high frequencies and conversely for negative asymmetries. This displaces the barycentre of the mode, such that neglecting the asymmetry by fitting with a symmetric profile, introduces a bias in the determination of the central frequency of the modes. Figure \ref{fig:Asym_PSF_Sun} shows that this effect, though small, is visible in the power spectrum of the Sun. To better visualise the asymmetry in the figure, the properties of the $l=0$ modes are isolated by dividing out the nearby $l=2$ mode and the noise background.
     Note that current studies of mode pulsations in Sun-like stars \citep[e.g.][]{Appourchaux2008, Benomar2012, Davies2015, Campante2016} considered that the asymmetry is negligible, which is equivalent to fixing $\chi_{n,l,m} = 0$. 
	 
\subsection{Assumption on the mode asymmetry} \label{sec:ass_asym}
\begin{figure}[t]
  \begin{center}
   
\subfigure{\epsfig{figure=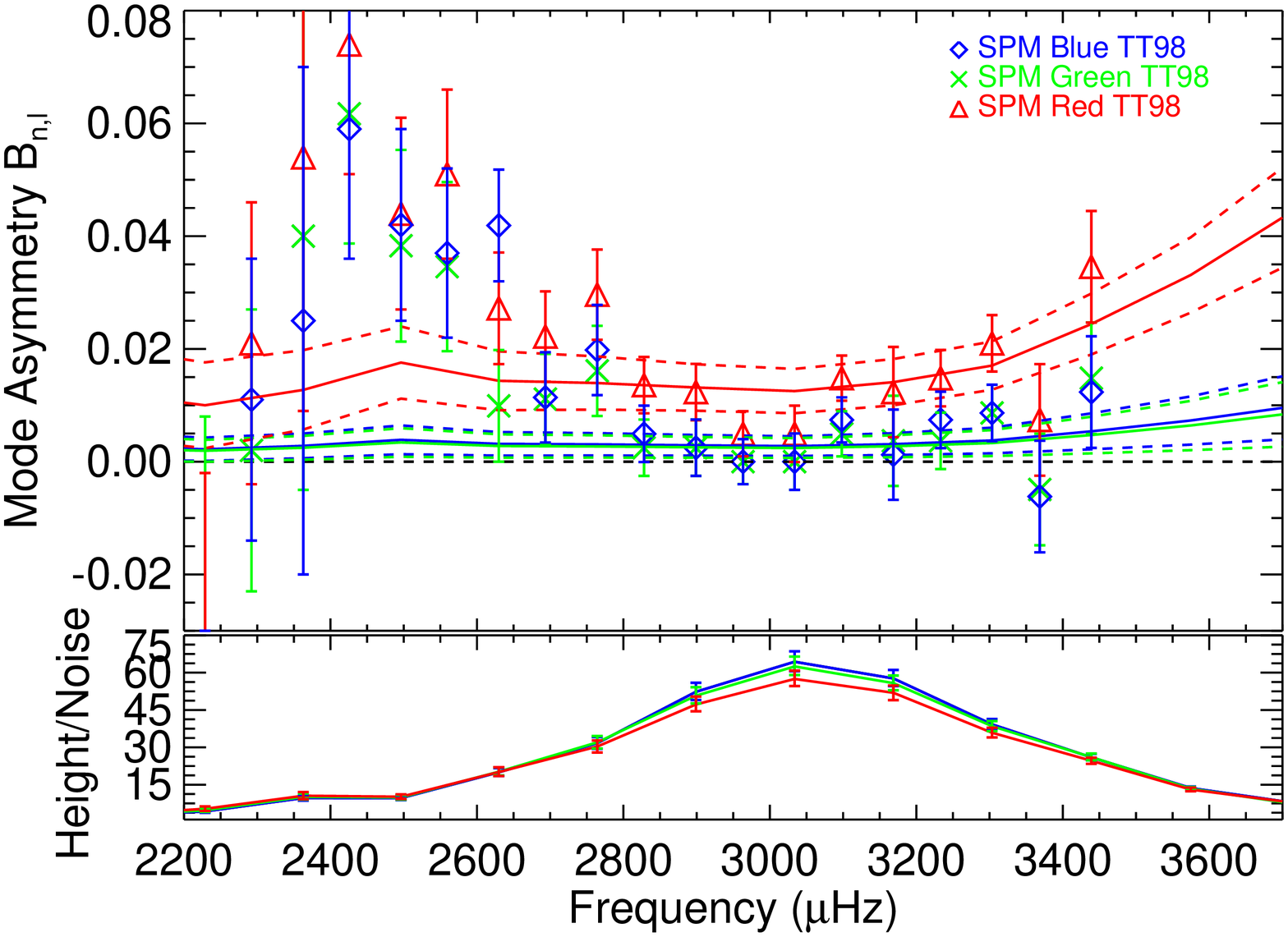, angle=0, width=8.8cm, height=6cm}}
  \end{center}
\caption{\textbf{Top.} Asymmetry coefficient $B_{n,l=0}$ and $B_{n,l=1}$ from TT98 
for the three Virgo/SPM \red{channels} (Blue, Green, Red). The best fit (solid line) from our analysis with a constant 
$\chi \propto B_{n,l}/\Gamma_{n,l}$. The $68 \%$ confidence interval is also shown (dashed lines). \textbf{Bottom.} Height-to-Noise for $l=0$ modes.}   
\label{fig:TT98_vs_me}
\end{figure}

This study focuses on determining the sign and the overall significance of the mode asymmetry for an ensemble of stars, which have signal-to-noise ratio three to 20 times lower to the Sun. Therefore, in order to improve the robustness of the measure\red{ment}, we need to simplify the description of the mode asymmetry. We opted for the simplest description which assumes 
that $\chi_{n,l,m} = \chi$ is constant over the range of observed modes. Figure \ref{fig:TT98_vs_me} compares the results from TT98 
with those from our analysis with a constant $\chi$. TT98 measured the mode asymmetry $B_{n,l=0}$ and $B_{n,l=1}$ using the SPM observations of May 1996, until December 1997 for the three channels (Red, Green, Blue) of the SPM/VIRGO instrument aboard the SoHo spacecraft. Similarly to {\it Kepler}, this instrument measures oscillations in intensity. The comparison is made possible by using equation \ref{eq:asym_chi} to express our results in terms of $B_{n,l}$. We found $\chi_{\mathrm{red}}=75.1 \pm 5.4$, $\chi_\mathrm{green}=12.1 \pm 4.9$ and $\chi_\mathrm{blue}=13.8 \pm 4.7$. The constant dimensionless asymmetry $\chi$ gives consistent asymmetry for modes with the largest height-to-noise ratio (HNR) and at high frequencies, for all the channels.
However, the asymmetry for modes with frequencies between $2400\,\mu$Hz and $2550\,\mu$Hz (ie, two or three radial orders out of eleven, depending on the channel) 
is apparently underestimated. Below $2200\,\mu$Hz, the large measurement errors given by TT98 
prevent us from evaluat\red{ing} the relevance of the constant-$\chi$ approximation. We conclude that the measure\red{ment} of the asymmetry is not accurate for low-frequency modes when assuming a constant coefficient $\chi$, but this remains sufficient to evaluate the importance of the mode asymmetry across the HR diagram. 

{\it Kepler} has a band pass that encompass{es} the three channels of the SPM instrument. Therefore to compare the solar asymmetry with {\it Kepler} measurements, we use the average asymmetry of the three SPM channels ($\chi_{\mathrm{sun}} = 33.7 \pm 1.7$).

\red{In velocity measurement, such as with the GOLF instrument aboard SoHo, the mode asymmetry is negative. Although there is no clear consensus on the origin of the sign reversal between photometry and velocity \citep[e.g.][]{Georgobiani2003, Nigam1998b}, it gives the opportunity to evaluate the capacity of our pulsation modeling to measure negative asymmetry coefficients.  We therefore fit\red{ted} the GOLF data (PM 1) over the same period as TT98 and found $\chi = - 26.6 \pm 1.0$. This corresponds to $B_{n,l} \simeq -0.01 \mu$Hz, which is comparable to what is found in the literature \citep[e.g.][]{Basu2000}. } 

	\subsection{Prior on the mode asymmetry $\chi$ for the MCMC analysis } \label{sec:priors:chi}
It is quite a difficult task to predict the sign and the value of the asymmetry because it requires a proper modeling of the properties 
of the excitation mechanisms. Therefore a non-informative prior on the absolute value of $\chi$ should be preferred. This is justified by the necessity to provide an unbiased measure of the parameters that could be correlated with the asymmetry. For example, it is established 
that neglecting the asymmetry of low-degree modes introduces a systematic error in the solar mode frequencies \red{that could be greater than the frequency precision.} 
\red{The induced systematic error on the inverted sound profile of the Sun's deep layers is likely negligible \citep{Basu2000}. For the Sun, this justifies the use of frequencies determined by neglecting the mode asymmetry. However, the internal structure and the fundamental parameters (mass, radius, age, etc.) of other stars is not determined by inversion, but rather by a forward modeling. This is because only low degree modes are available to us. Due to the lack of measurements of the asymmetry on other stars than the Sun, the influence of such systematic errors on stellar models is so far not known.} 

Since the mode asymmetry is a scale parameter, the most appropriate choice of prior is the (truncated) Jeffrey's non-informative prior \citep{Jeffreys1961},
\begin{align}
	p(\chi | M) = 
	\begin{cases}
		 \frac{C }{ |\chi| + \chi_0}  & \mbox{ if } 0 \leq | \chi | \leq \chi_{max} \nonumber \\
		0 & \mbox{ otherwise.}
	\end{cases}
\end{align}
Here, $C=1/\ln ( 1 + \chi_{max} / \chi_0 )$ is a normalisation constant. This prior is concentrated near zero so that in the absence of information about mode asymmetry in the data, the posterior distribution will be centered on zero and clearly depart from a Gaussian distribution. However, one needs to fix a value for $\chi_{0}$ which controls the degree of concentration near $0$ of the function, and $\chi_{max}$ which is required to ensure the integrability of the prior. Here we use $\chi_{0}=5$ and $\chi_{max}=200$. Fits for variable asymmetries for different modes ($\chi \approx \chi_{n,l}$) in the Sun do not exceed $\chi\approx100$ \citep{Toutain1998, Garcia2011b}, so this choice of $\chi_{max}$ is conservative. 

\section{Data selection and preparation} \label{sec:3}
In this work, the considered ensemble of stars is a subset of the \emph{Kepler} LEGACY sample \citep{Lund2017}. The LEGACY sample is made of stars that have been continuously observed for more than one year and with the highest signal-to-noise ratio. To measure the mode asymmetry reliably, it is preferable to focus on stars with unambiguous mode identification. This is possible when modes of degree $l=0$ and $l=2$ modes are well separated so that there is a limited crosstalk between the modes. A significant superposition of these modes, as seen predominantly in F stars \cite[e.g.][]{Appourchaux2008} reduces the precision on the mode properties. Due to this limit we selected 43 stars for analysis out of 66 of the LEGACY sample. The current analysis uses the un-weighted power spectra provided by the KASOC pipeline\footnote{\url{kasoc.phys.au.dk}} \citep{Handberg2014}. 

\section{Results} \label{sec:4}
\begin{figure}[t]
  \begin{center}
	\subfigure{\epsfig{figure=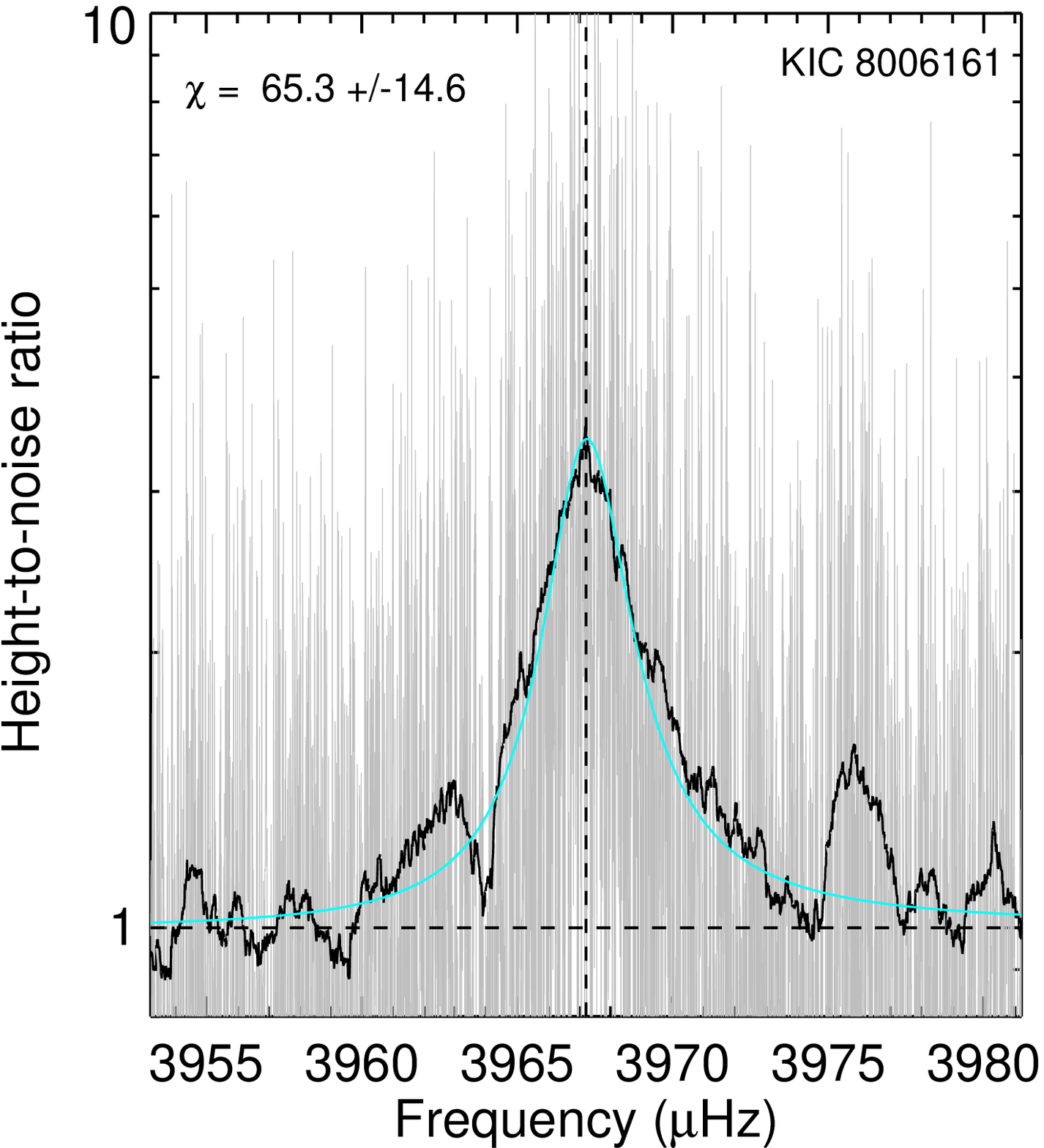, angle=0, width=5.8cm, height=6cm}} 
	\subfigure{\epsfig{figure=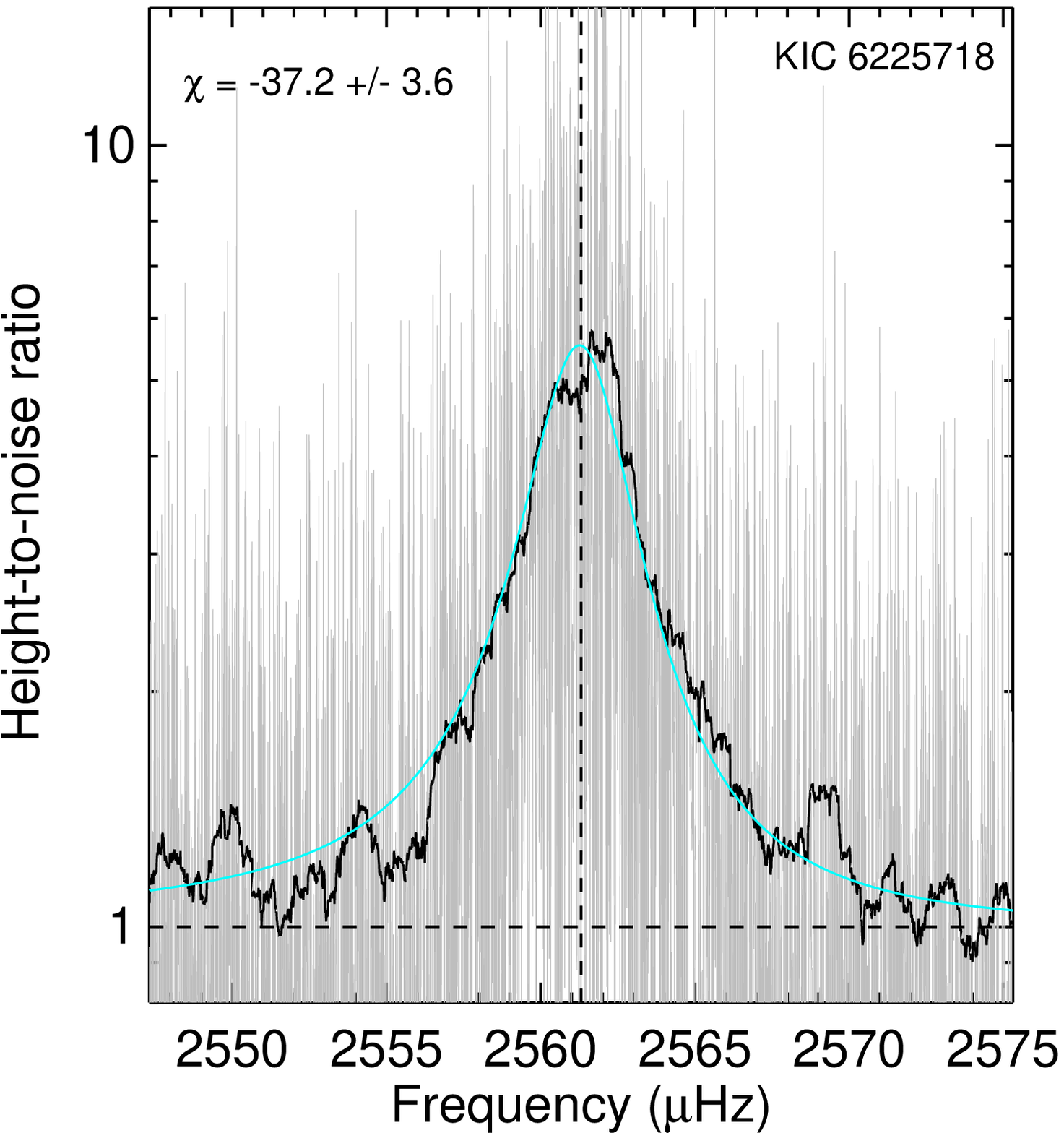, angle=0, width=6cm}} \\
  \end{center}
\caption{\textbf{Top}. Same as Figure \ref{fig:Asym_PSF_Sun}, but for KIC 8006161. Like the Sun, the asymmetry is positive so that power excess is visible at high frequency. \textbf{Bottom.} KIC 6225718 as an example of negative asymmetry, which results in a power excess at lower frequencies.}   
\label{fig:Asym_PSF}
\end{figure}

The main parameters from the fits of the 43 stars of our analysis are given in Table \ref{table:1}. This section discusses the main results of the analysis.
\red{Similar to the Sun, stars with the highest data quality (in terms of observation duration and signal-to-noise) have modes with clear asymmetry in the power spectrum. However, as illustrated in Figure \ref{fig:Asym_PSF} for KIC 8006161 and KIC 6225718, some stars have a positive asymmetry (elevated power in the Lorentzian tail toward high frequencies) in the power spectrum. Others show a negative asymmetry (elevated power in the tail toward lower frequency).}
	
	\subsection{Surface gravity and temperature dependence of the asymmetry} 
\begin{figure*}[t]
  \begin{center}
     \subfigure{\epsfig{figure=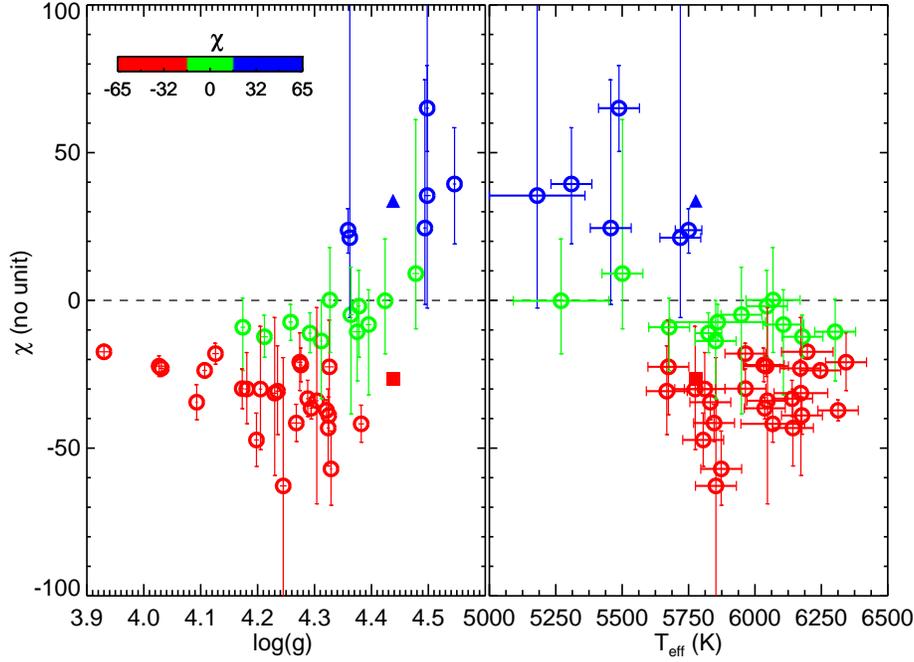, angle=0, width=12cm}} 
  \end{center}
\caption{\red{\textbf{Left.} Asymmetry coefficient $\chi$ as a function of the $\log(g)$. Colors indicate the importance of the asymmetry for {\it Kepler} stars (circle).  The asymmetry measured for the Sun is the filled triangle (photometry) and the filled square (velocity). \textbf{Right.} Asymmetry coefficient as a function of the effective temperature $T_{\mathrm{eff}}$.}}
\label{fig:Asym_vs_loggTeff}
\end{figure*}

\red{Figure \ref{fig:Asym_vs_loggTeff} shows that the mode asymmetry depends on the surface gravity $\log(g)$ and on the effective temperature $T_{\mathrm{eff}}$. In the figure, the $\log(g)$ values are determined by stellar modeling using AMP\footnote{Asteroseismic Modeling Portal, \url{https://amp.phys.au.dk/}} \citep{Metcalfe2009}. These are listed in \cite{Creevey2017} along with spectroscopic temperature measurements compiled for multiple studies. Stars cooler than $\simeq 5700K$ and with $\log(g)$ greater than $\simeq 4.4$ predominantly show positive asymmetry in the {\it Kepler} photometric data. Hotter and lower gravity stars show exclusively negative asymmetry.  The fact that the mode asymmetry depends on atmospheric properties of stars suggest that the mode asymmetry may change while the stars evolve.}

\subsection{Evolution of the asymmetry across the HR diagram}
\begin{figure}[t]
  \begin{center}
  	  	\subfigure{\epsfig{figure=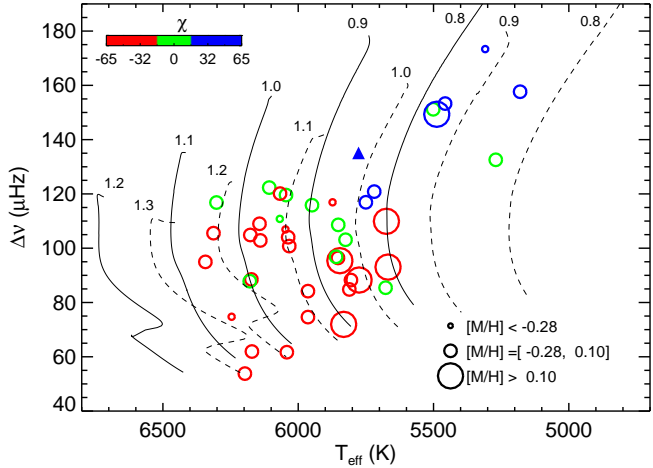, angle=0, width=8.55cm}}    
\end{center}
\caption{Seismic HR-diagram for the analysed stars, using the large separation $\Delta \nu$ and effective surface temperature $T_{\mathrm{eff}}$. Colors are for the asymmetry. Size of circles indicate the metallicity. Superimposed are evolution tracks for masses between 0.8 and 1.3 $M_{\mathrm{\odot}}$ for metallicity $\mathrm{[M/H]}=-0.21$ ($Z=0.008$, solid line) or $\mathrm{[M/H]}=0.10$ ($Z=0.016$, dashed line). The Sun is the triangle.} 
\label{fig:Asym_HRdiagram}
\end{figure} 
Figure \ref{fig:Asym_HRdiagram} shows the measured asymmetry coefficient, in the $\Delta \nu - T_{\rm eff}$ plane, where $\Delta \nu$ is the large separation and $T_{\rm eff}$ the effective temperature. As in \cite{White2011}, the figure is a modified HR-diagram where $\Delta\nu$ is used instead of the luminosity.
The stellar metallicity $[M/H]$ is also represented. Here, we use the spectroscopic temperature and metallicity of the LEGACY stars that have been compiled from various sources by \cite{Creevey2017}.
The large separation is determined by a linear fit of the measured radial mode ($l=0$) frequencies, listed in Table \ref{table:1}. For example in this diagram, KIC 8006161 has an effective temperature $T_\mathrm{eff}=5488 \pm 77$ K and $\Delta\nu=149.34 \pm 0.03\mu$Hz. Similarly, KIC 6225718 has $T_{\mathrm{eff}}=6313 \pm 76$ and $\Delta\nu=105.53\pm 0.04\mu$Hz.
The figure also shows evolution tracks from the ZAMS to the TAMS for stars with masses in the range $\mathrm{M}=[0.8, 1.3] \mathrm{M}_{\odot}$ and for two metallicities $Z=0.008$ and $Z=0.016$. These are representative of the observed surface metallicity of the sample, \red{which} has a non-Gaussian distribution, skewed toward low metallicity. This distribution is characterized by a median value 
$\overline{\mathrm{[M/H]}}=-0.05$, with most stars having metallicities\footnote{This range is computed using the confidence interval at $68\%$ centered on the median.} between $[M/H]=-0.28$ and $[M/H]=0.10$.

Here, the dispersion in [M/H] corresponds to the range $Z=[0.007, 0.016]$. 
The initial helium content is set to $0.27$. Adopting another value of the initial helium content  would only shifts the sequences horizontally (given the metallicity), reflecting the known mass-helium degeneracy \citep[see for instance][]{Lebreton2014}.  

Evolutionary sequences of stellar models were computed with the code CESTAM \citep{Marques2013} adapted from the original version CESAM2K \citep{Morel1997, Morel2008}. 
The input physics includes convection based on the \cite{Canuto1996} formalism (CGM) with a CGM mixing length parameter $\alpha_{GCM}=0.65$ from a solar calibration; the NACRE  nuclear reaction rates with the updated $^{14}N(p, \gamma)^{15}O$ reaction from \cite{Imbriani2005}. No convective core overshoot is added nor angular momentum transport. We adopted the AGSS09 solar mixture \citep{Asplund2009} with the relative mass fraction metal-to-hydrogen  abundance ratio $(Z/X)_\odot=0.0181$ and OPAL opacities \citep{Iglesias1996} completed with \cite{Alexander1994} opacities as well  as OPAL equation of state. Atomic diffusion  is included  following the formalism from \cite{Michaud1993}. Figure \ref{fig:Asym_HRdiagram} shows that low temperature stars have predominantly positive asymmetry, like the Sun. However, hotter stars have negative asymmetry. Observations of stars with $\Delta\nu \lesssim 120 \mu$Hz and with $T_{\mathrm{eff}} \lesssim 5500$K would be important to confirm this trend. Interestingly, stars with the smallest asymmetries occupy mostly a limited region that forms a band in the HR-diagram.
 \begin{figure}[t]
  \begin{center}
 
 \subfigure{\epsfig{figure=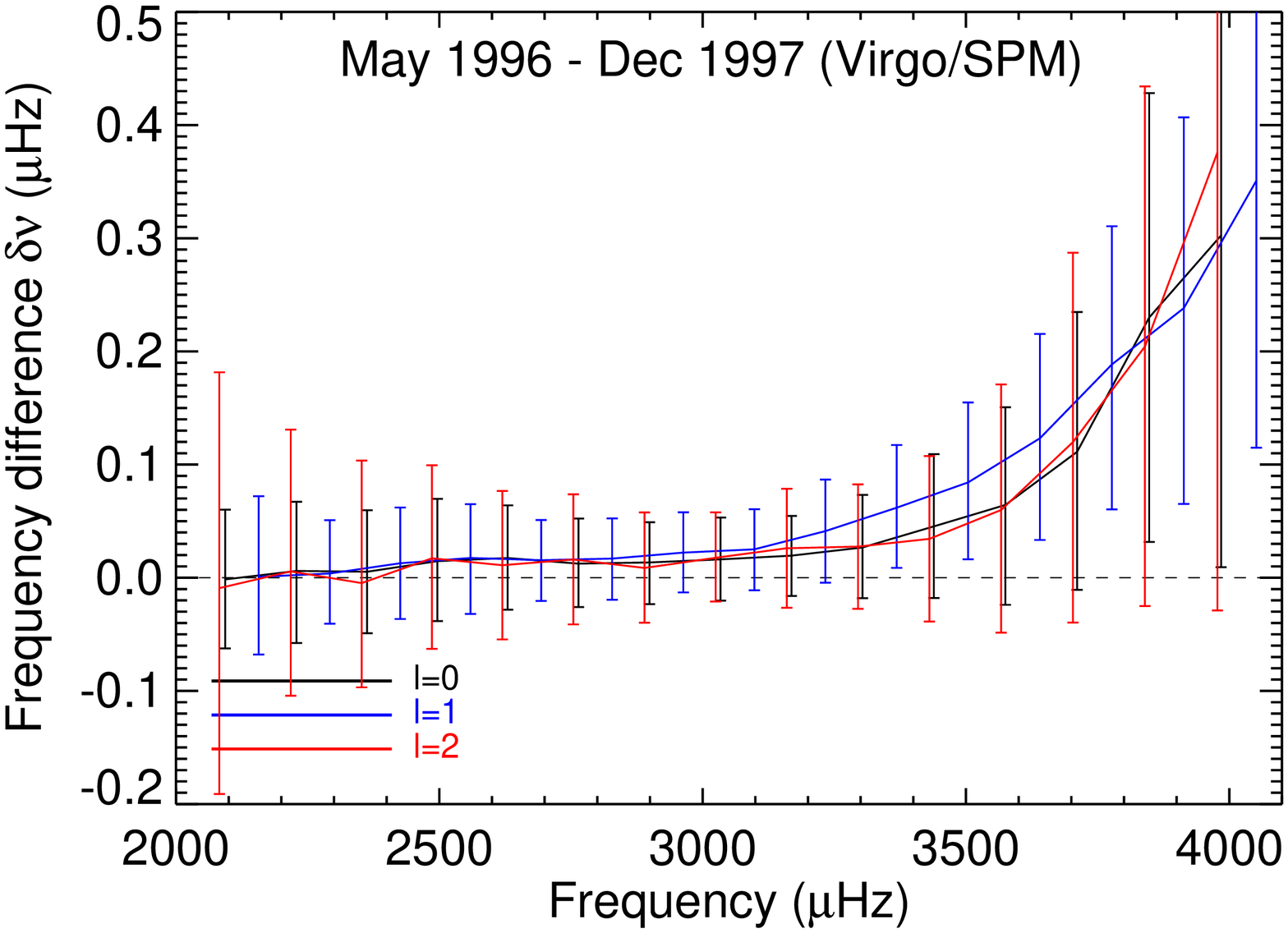, angle=0, width=7.1cm}}\\ 

\subfigure{\epsfig{figure=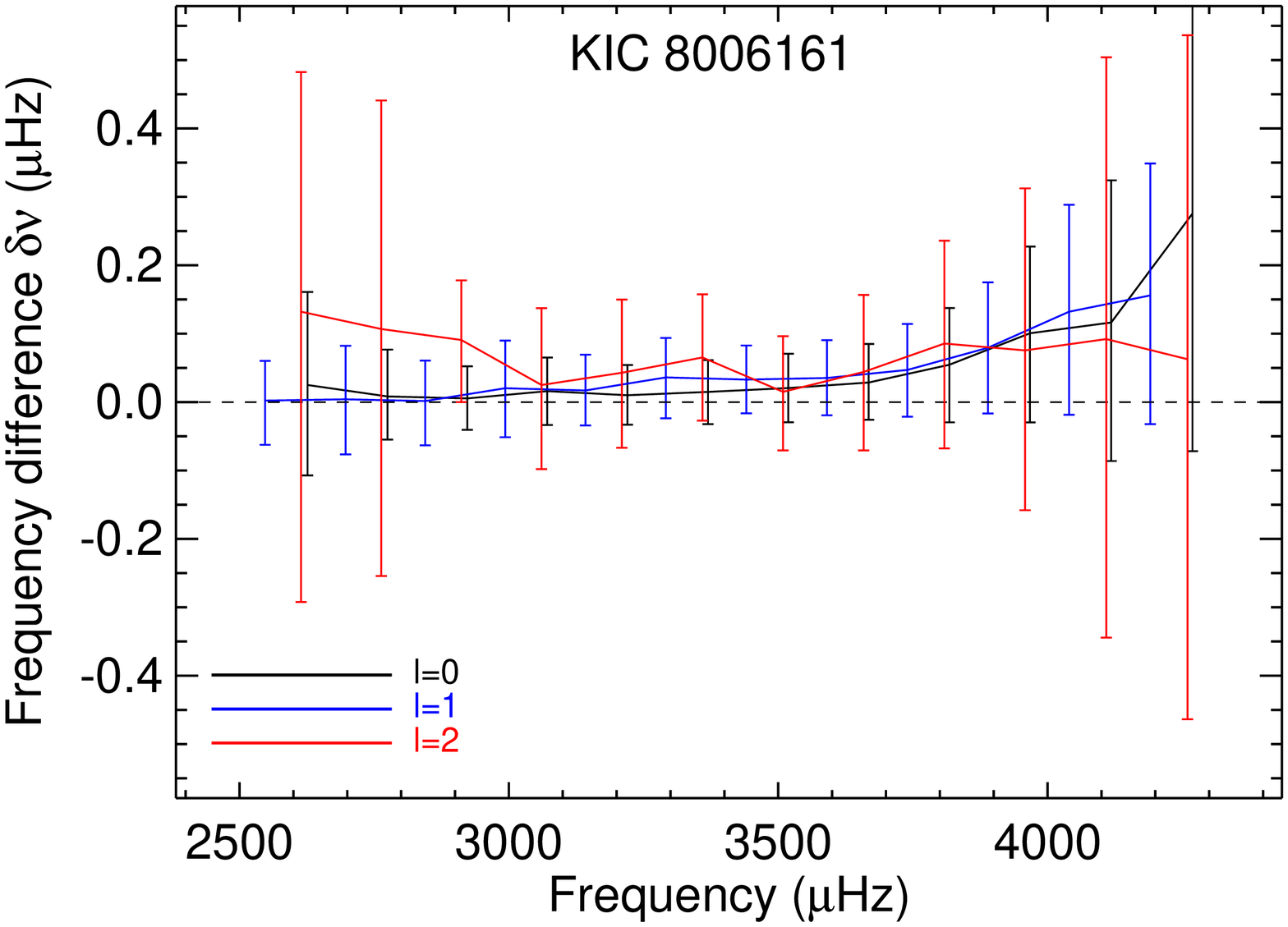, angle=0, width=7.1cm}}\\ 
	\subfigure{\epsfig{figure=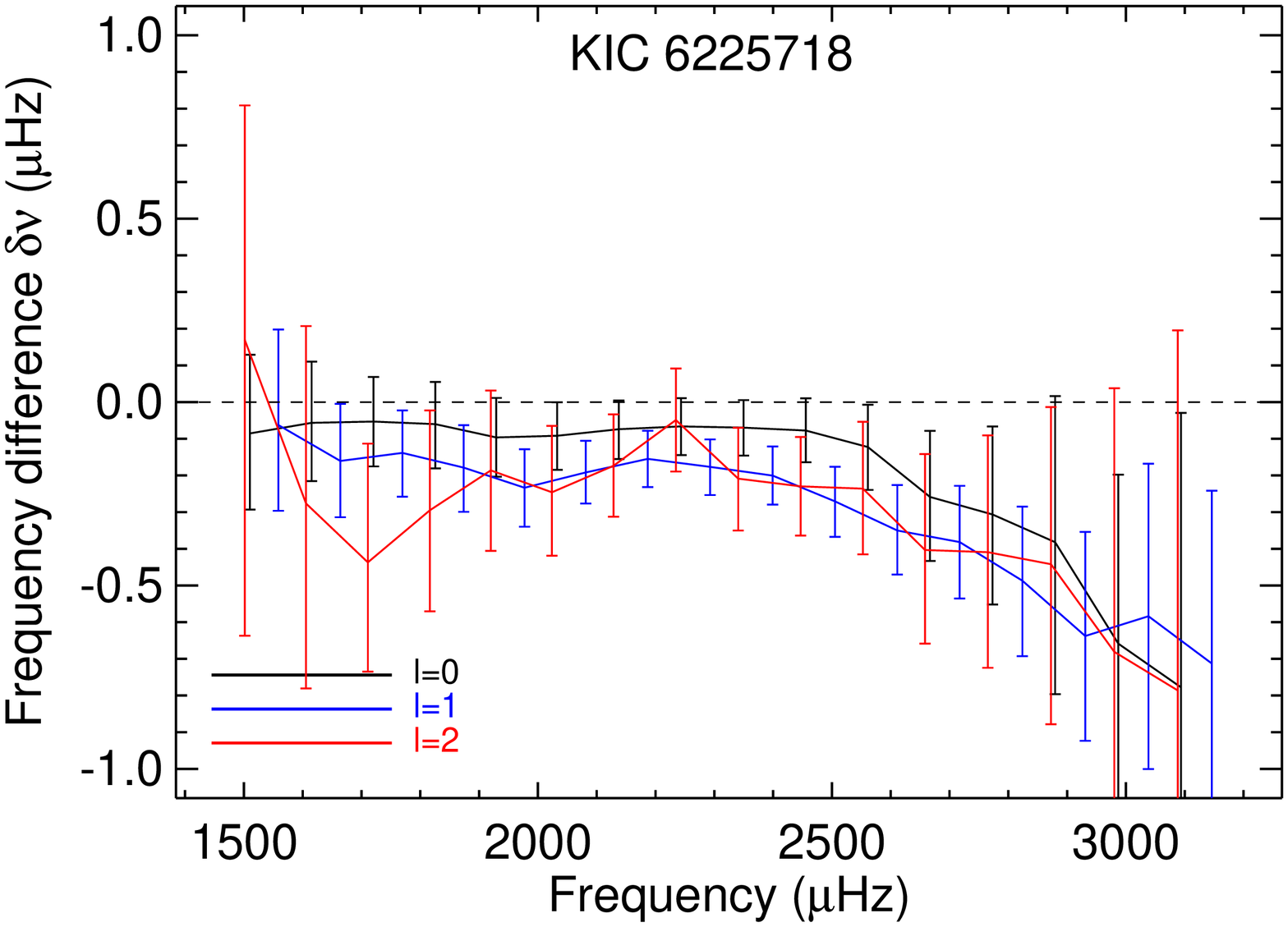, angle=0, width=7.1cm}} 
  \end{center}
\caption{Frequency difference $\delta\nu$ between a model with asymmetry and without asymmetry for the Sun with the SPM-green channel (top), KIC 8006161 (middle) and KIC 6225718 (bottom).}
\label{fig:Asym_vs_freq}
\end{figure} 

\subsection{Systematics on pulsation frequency}
Neglecting the mode asymmetry when fitting high quality-data such as those obtained by {\it Kepler} can have a significant impact on the measured central mode frequency. Figure \ref{fig:Asym_vs_freq} illustrates this by showing the frequency difference\red{, $\delta\nu = \nu_{n,l}^{\chi=0}-\nu_{n,l}^{\chi}$, induced by neglecting the mode asymmetry}, for the Sun (3 years of observations), KIC 8006161 and KIC 6225718. 
Here $\nu_{n,l}^{\chi=0}$ and $\nu_{n,l}^{\chi}$ are the measured frequencies when neglecting and including the mode asymmetry respectively.
When the asymmetry coefficient is positive ($\chi>0$), the central frequency of the mode is often overestimated by $1\sigma$ for the best observations (see e.g. the $l=1$ near the frequency at maximum amplitude of KIC 8006161). Conversely, the frequency is underestimated for negative asymmetry coefficients. As shown by Eq.\,\ref{eq:asym_chi}, the effective asymmetry $B_{n,l,m}$ is proportional to the mode width, so that the broader the mode, the greater the asymmetry. This translates into larger systematic errors at high frequencies, because the mode width increases \red{approximately} monotonically with frequency \citep[e.g.][]{Appourchaux2014}. This dependence is also directly observed in the power spectrum: modes with the most obvious asymmetry are the broadest.

The effect is less pronounced on cool stars because they tend to have narrower modes. Interestingly, the mode width of Sun-like stars has a plateau around the frequency at maximum amplitude $\nu_{max}$, so that the shift remains constant for modes with the highest signal-to-noise. Thus, it is evident that frequency differences such as the small separations \cite[e.g.][]{Roxburgh2003} are less affected by the mode asymmetry than the frequencies themselves. 
An important implication is that stellar modeling relying on frequency differences might be less biased than methods using absolute frequencies, when frequencies are measured assuming a symmetric Lorentzian.  One can argue that the biases introduced by neglecting the mode asymmetries are included in the ad-hoc corrections of surface effects. However, it leads to mix of the observational biases (due to the symmetric Lorentzian profile) together with theoretical uncertainties related to the poor description of the upper-most layers of stars. This is something to avoid because surface effects should be used as a seismic constraint to improve the physical description of the upper layers of stars (i.e. with a complete description of 3D convection processes near the surface). Consequently, it is fundamental to infer not only precise but also accurate observed frequencies. 

\begin{figure*}[t]
  \begin{center}
     \subfigure{\epsfig{figure=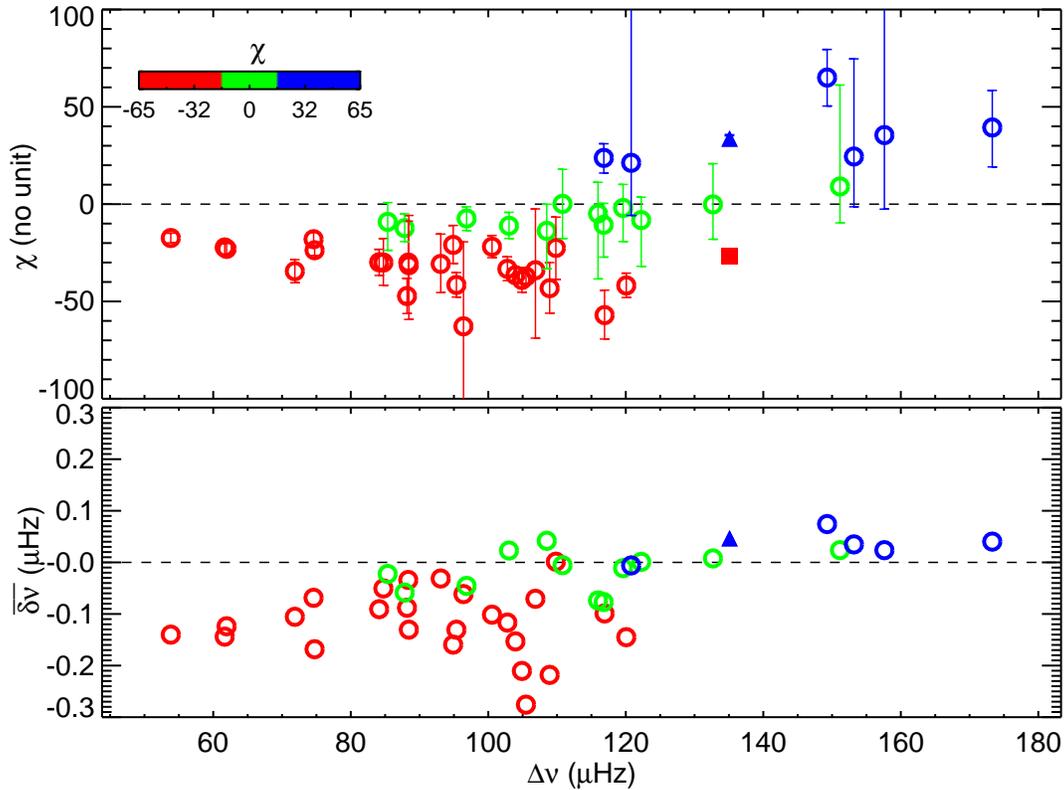, angle=0, width=14cm}} 
  \end{center}
\caption{\textbf{Top.} Asymmetry coefficient $\chi$ as a function of the large separation $\Delta\nu$. Below $\Delta\nu \approx 120$ $\mu$Hz, the asymmetry is mostly negative. For higher $\Delta\nu$ the asymmetry is positive like for the Sun in photometry (triangle). \red{The square shows the asymmetry coefficient of the Sun observed in velocity.} \textbf{Bottom.} Mean frequency difference $\overline{\delta\nu}$ 
between models with and without asymmetry.}
\label{fig:Asym_vs_Dnu}
\end{figure*}

Figure \ref{fig:Asym_vs_Dnu} shows the asymmetry coefficient $\chi$ and the average frequency shift $\overline{\delta\nu}$  as a function of the large separation $\Delta\nu$. 
The asymmetry slowly changes sign at $\Delta\nu \approx 120\,\mu$Hz. Below this value, $\chi \approx -30$, while for higher $\Delta\nu$ $\chi \approx +30$. 
As suggested by Figure \ref{fig:Asym_vs_freq}, neglecting the asymmetry leads to an overall frequency shift twice as large as for stars with negative asymmetry compared to those with positive asymmetry. This is due to the fact that the population with negative asymmetry is dominated by hotter stars, which tend to have larger mode widths.

\begin{table*}
\caption{Measured mode frequency ($\nu_{n,l}$), width ($\Gamma_{n,l}$), height ($H_{n,l}$), asymmetry ($\chi$) and their associated lower ($e_-$) and upper ($e_+$) uncertainties. The frequency shift $\delta\nu$ is also reported.}
\centering
\begin{tabular}{c|c|ccc|ccc|ccc|ccc|ccc}
KIC & $l$ & $\nu_{n,l}$ & $e_-$ & $e_+$ & $\Gamma_{n,l}$ & $e_-$ & $e_+$ & $H_{n,l}$ & $e_-$ & $e_+$ & $\chi$ & $e_-$ & $e_+$ & $\delta\nu$ & $e_-$ & $e_+$ \\ \hline
003427720  &    0 & 1969.7828     &     0.1858      &    0.1858   &   0.83  &    0.28   &   0.49   &   0.64  &    0.17  &    0.20  &  -2.0 &   17.3  &  12.1  &  0.0056 &   0.1793 &   0.1742 \\ 
003427720  &    0 & 2088.7509     &     0.3783      &    0.2904   &   1.32  &    0.35   &   0.36   &   0.52  &    0.11  &    0.16  &  -2.0  &  17.3  &  12.1  &  0.0691 &   0.2935 &   0.2537 \\
...       &  ...  &   ...         &     ...         &      ...    &  ..     &    ...    &   ...    &   ...   &    ...   &    ...   &   ...   &  ...   &  ...   &   ...   &    ...   &   ... \\
008006161  & 1  &    3739.7472        &  0.0692        &  0.0674     & 1.64    &  0.06   &   0.06   &   1.70   &   0.10   &   0.11  &  65.3  &  14.7  &  14.4  &  0.0470 &   0.0684  &  0.0672 \\
008006161  & 1  &     3889.3512   &       0.0984    &      0.0984   &   2.31  &    0.10  &    0.10   &   0.99  &    0.06   &   0.06 &   65.3  &  14.7  &  14.4  &  0.0794  &  0.0962 &   0.0956 \\
...       &  ...  &   ...         &     ...         &      ...    &  ..     &    ...    &   ...    &   ...   &    ...   &    ...   &   ...   &  ...   &  ...   &   ...   &    ...   &   ... \\
012258514   &  2 &  1998.6247    &      2.9698   &       2.4190  &    5.53    &  0.71  &    0.75   &   0.02   &   0.00  &    0.00 &  -18.0   &  3.6   &  3.5  & -0.0446   & 2.7716  &  2.3108 \\
\hline
\end{tabular} \label{table:1}
\tablecomments{Table \ref{table:1} is published in its entirety in machine-readable format. A portion is shown here for guidance regarding its form and content.}
\end{table*}

\section{Conclusion and discussion} \label{sec:5}
   Sun-like stars have stochastically excited modes whose properties (central frequency, width, amplitude) are usually measured by fitting symmetric Lorentzian profiles to the power spectrum. However, solar observations by \cite{Duvall1993} showed that mode profiles are better represented by an asymmetric Lorentzian. The origin of the asymmetry is commonly interpreted as mainly being due to the interference of waves produced at the source of the excitation with reflected waves arriving at the source once again \citep{Abrams1996}.
   
   
   We measured the mode asymmetry for 43 stars observed by {\it Kepler} and showed that it is not negligible in most of them. The mode asymmetry is found to gradually change sign as a function of the large separation. Actually, stars with similar effective temperature and mass as the Sun show predominantly a positive asymmetry, indicating that the modes have an excess of energy at high frequency. However, hotter and more massive stars have negative asymmetry (excess of power at low frequency). 

	The mode asymmetry significantly affects measurements of the central mode frequencies. The effect is small but exceeds $1\sigma$ uncertainties for the best observations. Particular care must therefore be taken when determining fundamental stellar parameters (mass, metallicity, etc.) when pulsation frequencies are measured assuming symmetric Lorentzian profiles. The parameter space is larger when considering the mode asymmetry, which may decrease the robustness of the fit. 
Neglecting the mode asymmetry could \red{therefore} be justified for short, low-quality observation, but not for long-term observations. Indeed, frequency shifts due to the mode asymmetry is measured even for one year {\it Kepler} time series. In long-term observations such as those provided by {\it Kepler}, the precision on frequencies derived with or without accounting for the constant mode asymmetry are the same. However, including the asymmetry improves the accuracy on the frequencies, so that it might be better to always account for it.

    
    The reversal of line profile asymmetry that we have measured is a potentially important diagnostic for the study of the driving mechanism of solar-like oscillations and thus for the study of convection in Sun-like stars. Improvements in models of p-mode line profiles will be required to fully understand the solar observations, and to extend the range of validity of the theory to stars with different effective temperatures and surface gravities.
    


\bibliographystyle{apj}

\clearpage

\end{document}